\documentclass[preprint,12pt]{elsarticle}

\usepackage[latin1]{inputenc}
\usepackage[T1]{fontenc}
\usepackage{graphicx}
\usepackage{amssymb}
\usepackage{amsfonts}
\usepackage{amstext}
\usepackage{amsbsy}
\usepackage{amsmath}
\usepackage{amsopn}
\usepackage{euscript}
\usepackage{bm} 
\usepackage{color}

\usepackage{hyperref}
\usepackage{threeparttable}

\usepackage{nomencl}
\makenomenclature
\usepackage{etoolbox}
\renewcommand\nomgroup[1]{%
  \item[\bfseries
  \ifstrequal{#1}{A}{Symbols}{%
  \ifstrequal{#1}{B}{Greek symbols}{%
  \ifstrequal{#1}{C}{Superscripts}{%
  \ifstrequal{#1}{D}{Subscripts}{%
  \ifstrequal{#1}{E}{Acronyms}{}}}}}%
]}

\begin{document}

\begin{frontmatter}

\title{INFLUENCE OF INITIAL CONDITIONS ON COHERENT STRUCTURES IN A ROUND JET}

\author[IMFT]{Philippe Reynier}

\author[IMFT]{Hieu Ha Minh}

\address[IMFT]{Institut de M\'ecanique des Fluides de Toulouse, Av. du Pr. Camille Soula, 31400 Toulouse, France}

\end{frontmatter}



\section{Introduction}

Coherent structures play a determinant role in several flows including the round jet. 
According to Lau \& Fisher (1975) the structure of the near-field of a round jet consists 
essentially in an serie of vortices moving downstream in the mixing layer of the jet. 
The regions located 
between the vortices are characterized by a high shearing which is at the source of the high 
level of turbulence in the shear-layer. This high level of turbulence leads to the spreading 
out of the jet. So, the coherent structures play  a fundamental role in the expansion process 
of the jet. This is confirmed by the numerical results of Verzicco \& Orlandi (1994). The large 
scale vortices originate from the shear-layer instability (e.g. Michalke 1984). The Kelvin-Helmholtz 
instability involves the roll-up of the shear-layer and the formation of vortex rings observed by 
Liepmann \& Gharib (1992) and simulated by Verzicco \& Orlandi (1994). According to the review of 
Ho \& Huerre (1984) the roll-up process is predominantly a two-dimensional phenomenon. The vortex 
rings evolving in the jet shear-layer grow by pairings which have been simulated by Grinstein et al 
(1987). In a laminar or transitional jet the pairing speeds up the transition to the 
fully-developed turbulence (e.g. Verzicco \& Orlandi 1994). 

But the behavior of the round jet depends on the initial conditions of the flow. 
According to Sahr \& G\"{o}kalp (1991) a jet with a laminar initially shear-layer has a larger capacity 
of entrainment than a turbulent one. As the spreading out of a jet is related to the presence of 
coherent structures, this shows that the behavior of the jet depends on the instability wich 
should occured. In homogeneous incompressible jets the main parameter to influence instability is the 
initial shear-layer thickness. According to Michalke (1984) for homogeneous jets the coherent structures 
are involved by the induction of the vorticity occuring in the shear-layer. For Cohen \& Wygnanski (1986) 
the increase of the initial shear-layer thickness reduces the number of unstable 
modes. Therefore, the main objective of this paper is to specify the influence of the inlet conditions 
on the presence of large scale coherent structures in the flow. Previous numerical investigations allowed 
the simulation of natural unsteadiness in coaxial jets (e.g. Reynier \& Ha Minh 1996) and 
in compressible round 
jets (e.g. Reynier 1995) using turbulence models. This method which is an alternative 
to the LES is chosen for this study to evaluate the influence of the initial conditions on the 
organized unsteadiness. Finally, the impact of the modelling on the predictions will be investigated 
using both the classical model proposed by Launder \& Sharma (1974) and the model recalibrated 
by Ha Minh \& Kourta (1993).

\section{Flow modelling}

The main objective of this paper is the evaluation of the influence of 
inlet conditions on the coherent unsteadiness in a quasi-incompressible round jet. In this 
flow, the three-dimensional effects are weak up to five diameters (e.g. Grinstein et al 1987). 
Moreover, three-dimensional instabilities are strongly coupled with the random turbulence 
which becomes predominantly at the end of the potential core located at four or five diameters 
(e.g. Sokolov et al 1981). 
In the near-field of an homogeneous jet, coherent structures originate from the Kelvin-Helmholtz 
instability. This last involves some fluctuations in the mixing layer which are at the source 
of the large scale vortices. According to Ho \& Huerre (1984), 
the roll-up of the shear-layer is predominantly a two-dimensional process. Therefore, 
two-dimensional simulations were executed for this study. As the code works for compressible flows, 
the mass-weighted average of Favre (1965) is used. The governing equations are the Navier-Stokes 
equations, a state equation for a perfect gas and the equations of the turbulence model.  As the 
flow to be computed is a round jet the numerical code used axisymmetric coordinates.

The flow pattern studied is characterized by high Reynolds numbers, the DNS and the LES cannot 
be applied. As a consequence the semi-deterministic modelling (e.g. Ha Minh \& Kourta 1993), 
a method close to the LES, has been chosen for this numerical study. 
This method allows the simulation of the coherent unsteadiness using turbulence models. 
The turbulence model used for the predictions is the $k-\epsilon$ model. Two versions of this model 
are used for the simulations in order to evaluate the influence of the constant $C_{\mu}$ on 
the simulation of the natural unsteadiness. 
Firstly, the model proposed by Launder \& Sharma (1974) with the usual set of constants has 
been chosen. The second model retained is the version proposed by Ha Minh \& Kourta (1993) 
with a set of constants recalibrated on a backward-facing step to take into 
account the coherent structures evolving in this flow. If this model has not been recalibrated 
on a round jet, the low value of the constant $C_{\mu}$ (which has for value $0.02$ in this model 
and $0.09$ in the version of Launder \& Sharma) should make easier the simulation of the natural 
instability. Indeed, a lower value of this constant should involve a less diffusive turbulence model.

\section{Computational aspects}

The numerical scheme used for the calculations is the finite volume method proposed 
by MacCormack (1981). This explicit-implicit algorithm uses the prediction-correction step 
technique and resolved the Navier-Stokes equations in a conservative form. 
The method is accurate to the second order in time and space. The numerical code has been 
already successfully applied to the simulation of natural unsteadiness in coaxial jets 
by Reynier \& Ha Minh (1996) and validated with the experimental data of Ribeiro (1972).

\begin{figure}
  \begin{center}
\includegraphics[width=80mm]{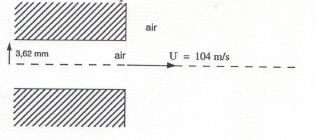}
  \caption{Configuration of the round jet}
    \end{center}
\end{figure}

An air jet presented in Figure 1 is computed. The 
exit velocity is 104 m/s, the Mach number of the jet is equal to 0.3 and the Reynolds 
number is 52240. The pipe diameter D is $7.24$ mm. Initially, the temperature at the inlet and in 
the whole domain is equal to $300^{o}$K, the pressure $P_{e}$ at the exit and in the 
computational field is $0.101$ MPa and the density in all the field is 
$\rho_{o}=1.28$ kg.m$^{-3}$.  The computational domain extends over 16.6 diameters in the 
streamwise direction and 8.3 diameters in the radial direction. The mesh uses 
100 $\times$ 93 cells with a coarse grid in the radial direction outside the jet. It is 
uniform in the streamwise direction.

\begin{figure}
  \begin{center}
\includegraphics[width=80mm]{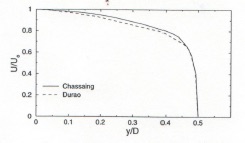}
  \caption{Profiles of velocity at the inlet}
    \end{center}
\end{figure}

\begin{figure}
  \begin{center}
\includegraphics[width=80mm]{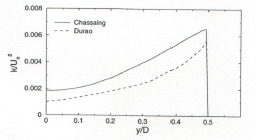}
  \caption{Profiles of turbulent kinetic energy at the inlet}
    \end{center}
\end{figure}

To investigate the influence of the inlet conditions on the flow 
unsteadiness several profiles of velocity (see Figure 2) and turbulent kinetic energy 
(see Figure 3) are applied at the inlet. The different computed cases are reported in the 
table 1. The initial conditions are derived from the experiment of Dur\~{a}o (1971) 
for the under-developed turbulence and from the measurements of Chassaing (1979) for the 
fully-developed turbulence.

\begin{table}[htb]
  \begin{center}
    \caption{Computed cases with different inlet conditions and turbulence models}
       \begin{tabular}{llll}
         \hline
         Mach number & Reynolds number & Inlet conditions & Turbulence model  \\ 
         \hline
         $0.3$ & $52240$ & Dur\~{a}o & Ha Minh \& Kourta  \\
         $0.3$ & $52240$ & Chassaing & Ha Minh \& Kourta  \\
         $0.3$ & $52240$ & Chassaing & Launder \& Sharma  \\
         \hline
       \end{tabular}
  \end{center}
\end{table}

Outside the jet, a wall is present (see Figure 1) in 
the transverse direction, so homogeneous Dirichlet conditions are applied on this boundary for 
velocity, turbulent kinetic energy and dissipation rate and homogeneous Neumann conditions are 
applied to density and pressure. The lower boundary  is the jet axis, therefore symmetry 
conditions are assumed. 
The upper boundary is located far from the flow then homogeneous Neumann conditions 
are imposed on this boundary. In order to do not perturb the flow, non-reflective conditions 
are applied at the outlet. They are deducted from characteristic relationships. They 
originated from the theory of the characteristic analysis and they have been developed for the Euler 
equations by Thompson (1987). When the flow is subsonic the pressure must be specified at the exit, 
an homogeneous Neumann condition is applied for this quantity. For the turbulent 
kinetic energy and its dissipation rate the same condition is used on this 
boundary.

\section{Results}

\subsection{Influence of inlet conditions on natural instability}

\begin{figure}
  \begin{center}
\includegraphics[width=80mm]{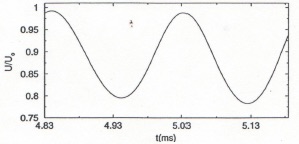}
  \caption{Time-variations of the streamwise velocity in the near-field at x$=$1.5D and y$=$0.5D}
    \end{center}
\end{figure}

\begin{figure}
  \begin{center}
\includegraphics[width=80mm]{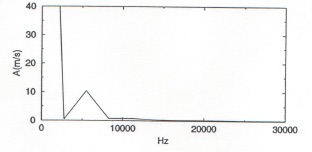}
 \caption{Spectrum of time-variations of the streamwise velocity in the near-field at 
x$=$1.5D and y$=$0.5D}
    \end{center}
\end{figure}

\begin{figure}
  \begin{center}
\includegraphics[width=120mm]{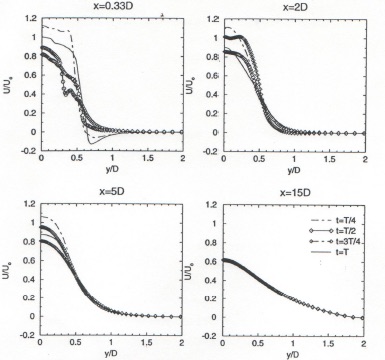}
 \caption{Profiles of the streamwise velocity for four sections located at x$=$0.33D, 
x$=$2D, x$=$5D and x$=$15D for four moments of a pseudo-period T/4, T/2, 3T/4 and T}
    \end{center}
\end{figure}

To evaluate the influence of initial conditions on the 
natural instability of the flow, the jet has been computed for fully-developed  
and under-developed turbulence at the inlet. If the velocity profiles (see Figure 2) are nearly 
similar it is not the case for the profiles of turbulent kinetic energy. The level of 
turbulence is lower for the under-developed case particularly in the central region of the flow 
where the level of turbulent kinetic energy is twice smaller that for the fully-developed turbulence. 
The turbulence model used for these simulations 
is the semi-deterministic model proposed by Ha Minh \& Kourta (1993). 
The executed computations with the inlet conditions derived from the experimental 
data of  Dur\~{a}o (1971) lead to the simulation of instabilities in the 
near-field without any flow excitation. The figure 4 represents the time-dependent 
variations of the streamwise velocity for a point located in the 
shear-layer at x$=$1.5D and y$=$0.5D. The variations of the streamwise velocity 
are quasi-sinusoidal. The corresponding spectrum obtained by Fourier 
analysis over one hundred periods are presented in figure 5. The spectrum puts in 
evidence the presence of a dominant frequency equal to 5600 Hz. The associated 
Strouhal number (calculate from the diameter of the inlet pipe and the exit 
velocity) is equal to 0.39. This Strouhal number is in the range of values contained between 
0.3 and 0.4 corresponding to the preferred mode (e.g. Michalke 1984). 
This preferred mode corresponds to the coherent structures which dominate the 
shear-layer of a round jet. In figure 6, the unsteady profiles of streamwise velocity are plotted. 
This figure shows the unsteady variations of the velocity for four sections of the mesh located 
at x$=$0.33D, x$=$2D, x$=$5D and x$=$15D, for four moments of a pseudo-period: 
T/4, T/2, 3T/4 and T. A high unsteadiness is active in 
the near region at x$=$0.33D and x$=$2D. This natural unsteadiness originates from 
the fluctuations in the shear-layer of the jet. The mixing layer becomes unstable 
near the inlet, due to the Kelvin-Helmholtz instability, then rolls up to form 
vortex rings. This phenomenon has been largely studied by Liepmann \& Gharib (1992) and 
Verzicco \& Orlandi (1994). The present results show the damping of 
the organized unsteadiness in the far field of the flow (see figure 6). At x$=$15D the 
unsteadiness has disappeared.

\begin{figure}
  \begin{center}
\includegraphics[width=120mm]{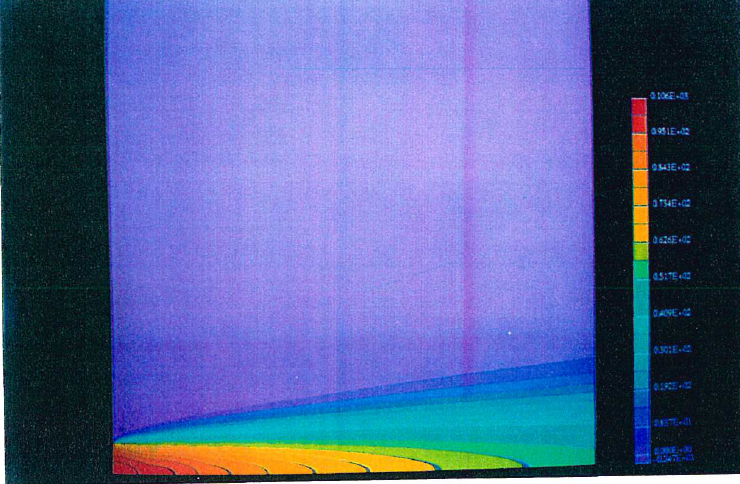}
 \caption{Streamwise velocity field predicts with the model of Ha Minh \& Kourta and a 
fully-developed turbulence at the inlet}
    \end{center}
\end{figure}

\begin{figure}
  \begin{center}
\includegraphics[width=120mm]{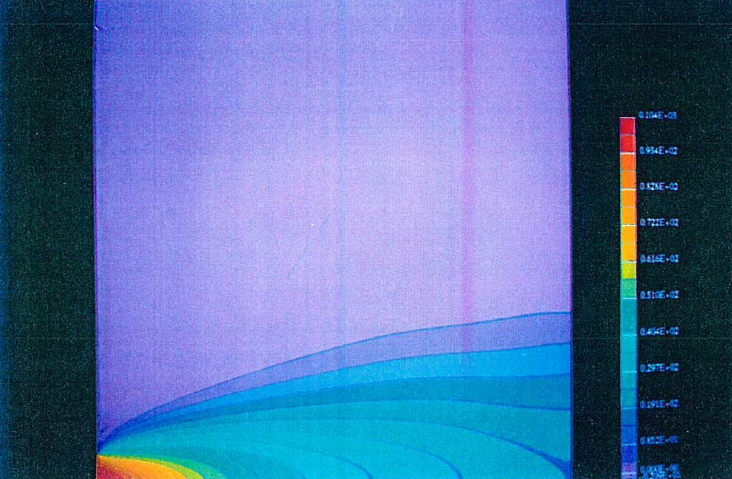}
 \caption{Radial velocity field predicts with the model of Ha Minh \& Kourta and a 
fully-developed turbulence at the inlet}
    \end{center}
\end{figure}

The simulation of the flow with a fully-developed turbulence at the inlet leads 
to the prediction of a steady field of streamwise velocity presented in figure 7. This lack of 
unsteadiness in the flow is due to a high turbulence level in the center of the 
jet at the inlet. According to Michalke (1984) the presence of coherent structures 
is involved by the induction of the vorticity occuring in the shear-layer. When the 
turbulence is fully-developed the vorticity occuring in the shear-layer is very weak or 
absent.  As a consequence the organized unsteadiness evolving in the shear-layer damps 
for a fully-developed turbulence at the inlet. The field of radial velocity 
predicted by the semi-deterministic model shows some residual instabilities in 
the shear-layer (see figure 8) but they are very weak. With the inlet conditions derived from the experiment 
of Dur\~{a}o (1971) the maximum of the radial velocity, scaled by the exit 
velocity, $|V| / U_{o}$, is equal to 0.33. With the inlet conditions corresponding 
to a fully-developed turbulence the maximum of this quantity is equal to 0.07. This low value of the 
radial velocity in the shear-layer shows a weak entrainment of the jet for the fully-developed turbulence 
that agrees with the experimental results of Sahr \& G\"{o}kalp (1991). The weak instabilities can be 
interpreted as remnants of coherent structures at a great age of turbulence.

\subsection{Influence of the modelling on the predictions}

\begin{figure}
  \begin{center}
\includegraphics[width=120mm]{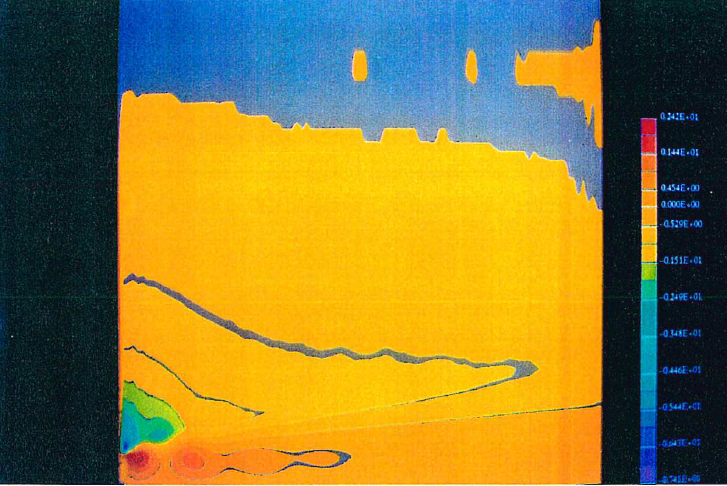}
\caption{Streamwise velocity field predicts with the model of Launder \& Sharma and a 
fully-developed turbulence at the inlet}
    \end{center}
\end{figure}

The flow has been computed with the initial conditions derived from the measurements of 
Chassaing (1979) for the two models retained for this study in order to evaluate the impact 
of the value of the constant $C_{\mu}$ on the predictions. The computations lead to the 
prediction of steady fields of streamwise velocity which are represented in figures 7 and 9. 
The visualizations show that the potential core is shorter for the simulation using the 
standard set of constants. With the classical 
model the expansion of the jet is larger than for the simulation with the recalibrated model. These 
discrepancies between the prediction of the two models are due to the difference of the value of the 
constant $C_{\mu}$. The low value of this constant in the model of Ha Minh \& Kourta (1993) 
involves a lower turbulent viscosity, therefore a less diffusive model. This explains 
the differences in the prediction of the streamwise velocity field. For the prediction 
of the radial velocity, the simulation using the classical model of Launder \& Sharma (1974) 
leads to steady results. But the results obtain using the 
semi-deterministic model (see figure 8) 
shows some weak unsteadiness in the shear-layer. This is a 
consequence of the smaller level of diffusion in the semi-deterministic model.

\section{Conclusion}

This numerical study shows that the coherent structures simulated with under-developed 
turbulence at the inlet are not found with fully-developed turbulence. Therefore the 
organized unsteadiness appears to be highly dependent on inlet conditions and particularly on the 
initial level of turbulent kinetic energy. Indeed, the coherent 
structures are characteristic of a young turbulence. The simulations with fully-developed 
initial conditions and the two turbulence models chosen for this study puts in evidence that 
a low value of the constant $C_{\mu}$ involves a weaker diffusion in the flow. This low diffusion 
allows the simulation of a weak unsteadiness in the shear-layer for a fully-developed turbulence at the 
inlet. These weak instabilities are interpreted as 
representative of remnants of coherent structures at a great age of turbulence.\\

\noindent
\textbf{Acknowledgements}

P. Reynier acknowledges with gratitude the support of the CNES through a post-doctorate grant.


\begin{thebibliography}{}

  \bibitem[Chassaing (1979)]{CH79}
    {\sc Chassaing, P.} (1979), 
    "M\'elange turbulent de gaz inertes dans un jet de tube libre", Ph. D. thesis,
    INPT n$^{o}$42, Toulouse.

  \bibitem[Cohen \& Wygnanski (1986)]{CW86}
    {\sc Cohen, J. \& Wygnanski, I.} (1986), 
    "The evolution of instabilities in the axisymmetric jet. Part 1. The linear 
     growth of disturbances near the nozzle", 
    {\em J. Fluid Mech.\ \/}{\bf 176}, 191--219.

  \bibitem[ Dur\~{a}o (1971)]{DU71}
    {\sc Dur\~{a}o, D.} (1971), 
    "Turbulent mixing of coaxial jets", Mast. of Science thesis,
    Imperial College of Science and Technology, London.

  \bibitem[Favre (1965)]{FA65}
    {\sc Favre, A.} (1965), 
    "Equations des gaz turbulents compressibles", 
    {\em J. de M\'ecanique\ \/} {\bf 4}, 361--390 and 391--421.

  \bibitem[Grinstein et al (1987)]{GOH87}
    {\sc Grinstein, F. F., Oran, E. S. \& Hussain, A. K. M. F.} (1987), 
    "Simulation of the transition region of axisymmetric free jets", 
    {\em 6$^{th}$ Symp. Turb. Shear Flows}, Toulouse.

  \bibitem[Ha Minh \& Kourta (1993)]{HK93}
    {\sc Ha Minh, H. \& Kourta, A.} (1993), 
    "Semi-deterministic turbulence modelling for flows dominated by strong 
    organized structures", 
    {\em 9$^{th}$ Symp. Turb. Shear Flows}, Kyoto.

  \bibitem[Ho \& Huerre (1984)]{HH84}
    {\sc Ho, C. H. \& Huerre, P.} (1984), 
    "Perturbed shear layer", 
    {\em Ann. Rev. Fluid Mech.\ \/} {\bf 16}, 365--424.

  \bibitem[Lau \& Fisher (1975)]{LF75}
    {\sc Lau, J. C. \& Fisher, M. J.} (1975), 
    "The vortex street structure of turbulent jets. Part 1", 
    {\em J. Fluid Mech.\ \/}{\bf 67}, 299--337.

  \bibitem[Launder \& Sharma (1974)]{LS74}
    {\sc Launder, B. E. \& Sharma, B. I.} (1974), 
    "Application of the energy dissipation model of turbulence to the calculation 
    of flow near a spinning disc", 
    {\em Lett. Heat Mass Transfer\ \/} {\bf 1}, 131--138.

  \bibitem[Liepmann \& Gharib (1992)]{LG92}
    {\sc Liepmann, D. \& Gharib, M.} (1992), 
    "The role of streamwise vorticity in the near field entrainment of round jets", 
    {\em J. Fluid Mech.\ \/}{\bf 245}, 643--668.

  \bibitem[MacCormack (1981)]{MC81}
    {\sc MacCormack, R.W.} (1981), 
    "A numerical method for solving the equations of compressible viscous flow", 
    {\em AIAA Paper\ \/}, 81-0110.

  \bibitem[Michalke (1984)]{MI84}
    {\sc Michalke, A.} (1984), 
    "Survey on jet instability theory", 
    {\em Prog. Aerospace Sci.\ \/} {\bf 21}, 159--199.

  \bibitem[Reynier (1995)]{RE95}
    {\sc Reynier, P.} (1995), 
    "Analyse physique, mod\'elisation et simulation num\'erique des jets simples 
    et des jets coaxiaux turbulents, compressibles et instationnaires", 
    Ph. D. thesis, INPT n$^{o}$1062, Toulouse.

  \bibitem[Reynier \& Ha Minh (1996)]{RH96}
    {\sc Reynier, P. \& Ha Minh, H.} (1996), 
    "Numerical prediction of unsteady compressible turbulent coaxial jets", 
   {\em Computers and Fluids\ \/}, accepted.

  \bibitem[Ribeiro (1972)]{RI72}
    {\sc Ribeiro, M. M.} (1972), 
    "Turbulent mixing of coaxial jets", Mast. of Science thesis,
    Imperial College of Science and Technology, London.

  \bibitem[Sahr \& G\"{o}kalp (1991)]{SG91}
    {\sc Sahr, B. \& G\"{o}kalp, I.} (1991), 
    "Variable density effects on the mixing of turbulent rectangular jets", 
   {\em 8$^{th}$ Symp. Turb. Shear Flows}, Munich.

  \bibitem[Sokolov et al (1981)]{SKH81}
    {\sc Sokolov, M., Kleis, S. J. \& Hussain, A. K. M. F.} (1981), 
    "Coherent structures induced by two simultaneous sparks in an axisymmetric jet", 
    {\em AIAA J.\ \/}{\bf 19}, 1000--1008.

  \bibitem[Thompson (1987)]{TH87}
    {\sc Thompson, K. W.} (1987), 
    "Time dependent boundary conditions for hyperbolic systems", 
    {\em J. Computational Physics\ \/}{\bf 68}, 1--24.

  \bibitem[Verzicco \& Orlandi (1994)]{VO94}
    {\sc Verzicco, R. \& Orlandi, P.} (1994), 
    "Direct simulations of the transitional regime of a circular jet", 
    {\em Phys.\ Fluids\ \/}{\bf 6}, 751--759.

\end {thebibliography}
\end{document}